\documentclass[sigconf]{acmart}
\usepackage[ruled,linesnumbered]{algorithm2e}
\usepackage{enumitem}
\usepackage{booktabs}
\usepackage{xcolor}
\usepackage{caption}
\usepackage{subcaption}
\usepackage{amsmath}
\usepackage{threeparttable}
\newcommand{\nop}[1]{}
\AtBeginDocument{%
  \providecommand\BibTeX{{%
    \normalfont B\kern-0.5em{\scshape i\kern-0.25em b}\kern-0.8em\TeX}}}

\copyrightyear{2021} 
\acmYear{2021} 
\setcopyright{acmlicensed}\acmConference[SIGIR '21]{Proceedings of the 44th International ACM SIGIR Conference on Research and Development in Information Retrieval}{July 11--15, 2021}{Virtual Event, Canada}
\acmBooktitle{Proceedings of the 44th International ACM SIGIR Conference on Research and Development in Information Retrieval (SIGIR '21), July 11--15, 2021, Virtual Event, Canada}
\acmPrice{15.00}
\acmDOI{10.1145/3404835.3462934}
\acmISBN{978-1-4503-8037-9/21/07}

\begin{document}
\fancyhead{}

\title{DyDiff-VAE: A Dynamic Variational Framework for Information Diffusion Prediction}

\author{Ruijie Wang$^{1}$, Zijie Huang$^{2}$, Shengzhong Liu$^{1}$, Huajie Shao$^{1}$, Dongxin Liu$^{1}$} 
\author{Jinyang Li$^{1}$, Tianshi Wang$^{1}$, Dachun Sun$^{1}$, Shuochao Yao$^{3}$ and Tarek Abdelzaher$^{1}$}

\affiliation{%
  \institution{$^{1}$ Department of Computer Science, University of Illinois at Urbana Champaign}
}

\affiliation{%
  \institution{$^{2}$ Department of Computer Science, University of California, Los Angeles}
}

\affiliation{%
  \institution{$^{3}$ Department of Computer Science, George Mason University}
}

\email{{ruijiew2, sl29, hshao5, dongxin3, jinyang7, tianshi3, dsun18, zaher}@illinois.edu}
\email{zijiehuang@cs.ucla.edu, shuochao@gmu.edu}

\renewcommand{\shortauthors}{Wang, et al.}

\begin{abstract}
This paper describes a novel diffusion model, DyDiff-VAE, for information diffusion prediction on social media. Given the initial content and a sequence of forwarding users, DyDiff-VAE aims to estimate the propagation likelihood for other potential users and predict the corresponding user rankings. Inferring user interests from diffusion data lies the foundation of diffusion prediction, because users often forward the information in which they are interested or the information from those who share similar interests. Their interests also evolve over time as the result of the dynamic social influence from neighbors and the time-sensitive information gained inside/outside the social media. Existing works fail to model users’ intrinsic interests from the diffusion data and assume user interests remain static along the time. DyDiff-VAE advances the state of the art in two directions: (i) We propose a dynamic encoder to infer the evolution of user interests from observed diffusion data. (ii) We propose a dual attentive decoder to estimate the propagation likelihood by integrating information from both the initial cascade content and the forwarding user sequence. Extensive experiments on four real-world datasets from Twitter and Youtube demonstrate the advantages of the proposed model; we show that it achieves \textbf{$43.3\%$} relative gains over the best baseline on average. Moreover, it has the lowest run-time compared with recurrent neural network based models.

\end{abstract}


\maketitle
\section{introduction}
\label{sec:intro}

\begin{figure*}[t]
    \centering
    \includegraphics[width = 1.0 \linewidth]{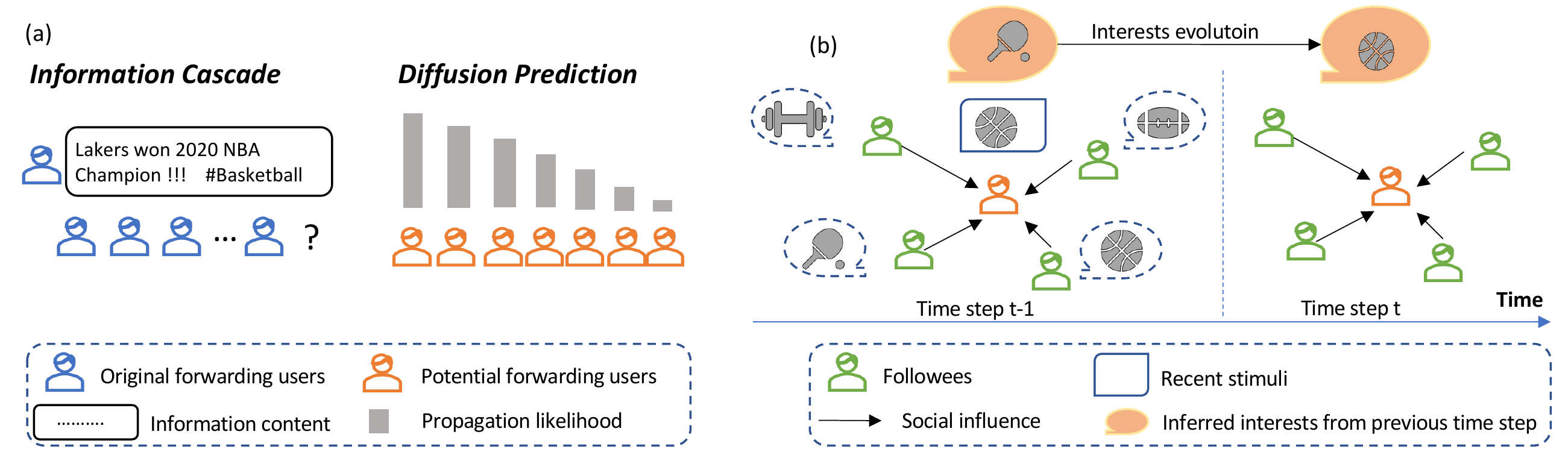}
    \caption{(a): An illustration of diffusion prediction task, which estimates the propagation likelihood for potential users and ranks them accordingly. (b): An illustration of dynamic user interests, which evolve over time driven by both social influence and recent stimuli.}
    \label{fig:illustration}
\end{figure*}

The paper advances the state of the arts of information diffusion prediction, a topic that receives much recent attention in several contexts, including social recommendations~\cite{GraphRec,hashtag,Github}, misinformation detection~\cite{misinfo}, polarization analysis~\cite{polarization}, user personalization~\cite{user1,user2,user3}, among others. On social media, information is propagated rapidly through users' posting and forwarding behaviors, leading to information cascades consisting of the users who have forwarded the content (we call them \textit{original forwarding users}) as well as the actual \textit{information content}. Given an information cascade, the diffusion model aims to estimate the propagation likelihood for other potential users and predict the user rankings, as shown in Figure~\ref{fig:illustration}a. 

On social media, users usually forward a piece of information out of their interests: On one hand, because of the widespread deployments of recommendation systems and search engines on social platforms, users are more likely to be exposed to and thereby propagate the contents matching their interests. On the other hand, users would also tendentiously pay attention to and get infected by the information from social neighbors, as a result of their shared interests and frequent interactions~\cite{homophily}. Therefore, how to infer user interests lies the foundation for diffusion prediction problems. However, it is non-trivial because user interests cannot be observed directly from their profiles due to various user privacy protection mechanisms, and they are highly dynamic over time. Existing neural diffusion models~\cite{NDM,TopoLSTM,DeepDiffuse,Inf-VAE}, despite demonstrating significant improvements over traditional methods built upon pre-defined propagation hypotheses~\cite{IC,LT,embIC,timeIC}, have two main drawbacks that oversimplify the inference of user interests. Firstly, they model user interests implicitly: they initialize user representations as random variables and optimize them by minimizing the distance between representations of cascades and forwarding users, with the objective that the learned user representations would be good approximations of user interests. However, user interests can be inferred explicitly via the diffusion data, e.g., cascade contents that users has previously forwarded. In other words, instead of building a discriminative model that learns user interests from scratch, we propose to build a generative model where user interests is modeled explicitly and serves as the latent variable that drives the diffusion process forward.

Moreover, existing works assume user representations remain static despite that user interests evolve over time in real world. Such dynamics are mainly driven by two factors: (i) \textit{social influence} received from their neighbors on the social networks (e.g., followees on Twitter), as individuals are likely to be influenced by their friends~\cite{co-evolve}. For instance, as shown in Figure~\ref{fig:illustration}b, a Twitter user is inferred to be interested in basketball-related topics at time step $t$ because, in part, that one of the followees likes it at time step $t-1$ . (ii) \textit{recent stimuli} which refer to information that users gained inside/outside the social media recently. Users can gain new information by reading news, forwarding posts online and etc, which also shapes the evolution of user interests. In Figure~\ref{fig:illustration}b, another reason why the Twitter user is inferred to be interested in NBA is that (s)he reads some related tweets recently. Modeling the evolution of user interests is critical for predicting propagation behaviors, as the relevance between information cascades and potential users varies as evolution of user interests. 

In this paper, we propose a dynamic generative model following the framework of variational autoencoder (VAE): DyDiff-VAE, which advances the existing diffusion models in two directions. First, we design a dynamic encoder to infer latent user interests over time by considering both social influence and recent stimuli as driven factors. We embed the graph convolutional layer~\cite{gcn} aggregating social influence from neighbors into the gated recurrent unit (GRU)~\cite{GRU}, which takes recent stimuli as inputs, to learn the recurrent function of interest evolution. Second, we propose a dual attentive decoder to estimate the propagation likelihood and corresponding rankings based on the inferred latent interests. The proposed dual attention mechanism, as a replacement of the recurrent neural networks (RNNs) used in other state-of-the-arts~\cite{TopoLSTM,DeepDiffuse,SNIDSA,CYANRNN}, integrates the heterogeneous information from both the user sequence and the cascade content, and overcomes back-propagation through time through efficient parallelizable attentions. Thus, it learns better representations of the information cascades and achieve higher efficiency. We conduct extensive experiments on four real-world datasets collected from the Twitter and Youtube to evaluate the performance of the proposed DyDiff-VAE. The evaluation results demonstrate that DyDiff-VAE outperforms other compared methods including diffusion models~\cite{DeepDiffuse,SNIDSA,TopoLSTM,Inf-VAE}, recommendation models~\cite{GraphRec} and graph learning methods (dynamic/static)~\cite{node2vec,gae,VGRNN}. It achieves \textbf{$43.3\%$} relative gains on average over the best baseline on four datasets. And it demonstrates the best efficiency compared to the RNN-based diffusion models. 

The rest of the paper is organized as follows: Section~\ref{sec:related} summarizes the related works, followed by the preliminaries in Section~\ref{sec:pre}. In Section~\ref{sec:model}, we introduce the proposed framework in detail. Section~\ref{sec:experiment} evaluates its performance on real-world datasets. Finally, we conclude the paper in Section~\ref{sec:conclusion}.

\section{related work}
\label{sec:related}

\textbf{Diffusion prediction.}
Early work for diffusion prediction usually holds certain assumptions about the diffusion process. For example, Independent Cascade (IC) models~\cite{IC,LT} popularized the study of information diffusion with a pairwise independence assumption. By introducing stronger assumptions about time-delay information, extensions of these models~\cite{continuousIC,CONNIE,NetInf,Netrate,Infopath,MMRate} considered temporal evolution in the diffusion process. However, pairwise independence oversimplifies the complex nature of the diffusion process, resulting in poor performance on real-world datasets. With advances in representation learning and deep learning, recent works~\cite{DeepDiffuse,TopoLSTM,CYANRNN,SNIDSA,infoRL,HDAN,Inf-VAE} design end-to-end frameworks to automatically learn diffusion patterns from data. They project cascades into the user-characterized space and predict the diffusion likelihood based on the distance between cascade and user representations. Most of these models~\cite{DeepDiffuse,TopoLSTM,CYANRNN,SNIDSA,infoRL} employs extensions of Recurrent Neural Networks (RNNs) to extract meaningful features from user sequences, e.g., local topology among users and temporal patterns within the diffusion cascades. Recent work~\cite{Inf-VAE} enhances user representations, instead of initializing randomly, via a variational graph autoencoder. However, all of the above neural methods focus entirely on learning the cascade representations from user sequence, ignoring information content which has been shown meaningful in information diffusion~\cite{DeepText}. Moreover, existing works ignore modeling dynamic user interests from diffusion data. To the best of our knowledge, DyDiff-VAE is the first framework that jointly infers the dynamic user interests and models information content for diffusion prediction.

\noindent
\textbf{Variational autoencoder.}
Variational autoencoder (VAE)~\cite{VAE} models the generative process of data $\mathbf{x}$ from the latent variable $\mathbf{z}$. The encoder approximates the distribution of latent variable $\mathbf{z}$ by the variational posterior, and the decoder generates the data $\mathbf{x}$ given $\mathbf{z}$. VAE has superior performance in representation learning for both static data~\cite{gae,betaVAE,controlVAE} and dynamic data~\cite{VGRNN,LG-ODE}. The key difference between DyDiff-VAE and existing VAE on modeling dynamic data is the design of the decoder. The conventional decoder aims to reconstruct the data samples or the pair-wise relevance among data samples, while our dual attentive decoder aims to reconstruct the propagation likelihood between the information cascade and potential users. It explicitly models the sequential and heterogeneous information via the dual attentive decoder from both the user sequence and the information content.  

\noindent
\textbf{Sequence modeling.}
Representing the information cascade requires an efficient sequence modeling of long user sequence. Recurrent Neural Networks (RNNs) are one of the most popular sequence models, which model the cross-dependency within a sequence by a special design of various gate mechanisms. Several RNN variants are proposed for handling different sequence types, (e.g., GRU and LSTM~\cite{GRU,LSTM} for long sequence modeling where the gradient vanishing and exploding problems are addressed, tree-LSTM~\cite{treeLSTM} for natural language with a latent grammar structure, etc). However, RNNs scale poorly with the increasing in sequence length due to the sequential nature of back-propagation through time, which becomes the bottleneck for large-scale applications. To improve computational efficiency, alternative transformation methods that include a parallel convolutional operator~\cite{conSeq} and self-attention~\cite{DySAT,transformer} have been proposed. Existing diffusion models usually utilize RNNs, resulting in prohibitive costs for long cascade sequences. To achieve better efficiency, we propose an efficient and parallelizable dual attentive decoder to learn cascade representations.

\section{preliminaries}
\label{sec:pre}

In this section, we introduce the definitions and the problem formulation in this paper. We split time span into discrete time steps, across which user interests are evolving driven by recent stimuli and social influence. According to ~\cite{short}, over $99\%$ information cascade usually lasts shorter than one day,  while the user interests evolve much more slowly. Thus, We assume the information cascade happens fully within one time step. We define recent stimuli, social influence, and information cascades as follows: 

\begin{definition}[\textbf{Recent stimuli}]
Recent stimuli refer to the cascade contents propagated by users at the last time step. Through the propagation behaviors, users gain stimuli inside the social media to change their interests, which in turn affects their behaviors in the following time steps. 
\end{definition}

\begin{definition}[\textbf{Social Influence}]
Social influence refers to factors describing how users' social neighbors influence their interests through social relations, (e.g., follower-followee relations on Twitter), which are reflected by social networks. We represent the static social network as $\mathcal{G} = (\mathcal{U},\mathcal{E})$, where $\mathcal{U}$ is the set of users, and $\mathcal{E}$ is the set of asymmetrical social relations among users.
\end{definition}

\begin{definition}[\textbf{Information Cascades}]
Let $I^{(t)}_j$ denotes the $j$-th information cascade happening at time step $t$. It is defined as a tuple of information content $c_j$ and a diffusion sequence of original forwarding users in ascending order of participation time. Thus, $I^{(t)}_j = (c_j, \{u_{k}\}_{k = 1} ^{K})$, where $K$ denotes the length of the user sequence. We denote the set of information cascade starting at time step $t$ as $\mathbb{I}^{(t)}$.
\end{definition}

Based on the definitions above, we formulate the diffusion prediction problem as follows:

\begin{definition}[\textbf{Diffusion Prediction}]
Given an information cascade $I^{(t)}_j = (c_j, \{u_{k}\}_{k = 1} ^{K})$ at time step $t$, the diffusion prediction aims to estimate the diffusion likelihood $p(u_i|I_j^{(t)})$ for potential user $u_i\in \mathcal{U} \setminus \{u_{k}\}_{k = 1} ^{K}$ by utilizing the information gained from all observed cascades $\bigcup_{n=1}^{n<t}\mathbb{I}^{(n)}$ before $t$ and the social network $\mathcal{G}$. The outcome is the propagation likelihood for the potential forwarding users as well as the corresponding rankings.

\end{definition}

\section{methodology}
\label{sec:model}
In this section, we describe the framework of DyDiff-VAE in detail.

\begin{figure*}[t]
    \centering
    \includegraphics[width = 1. \linewidth]{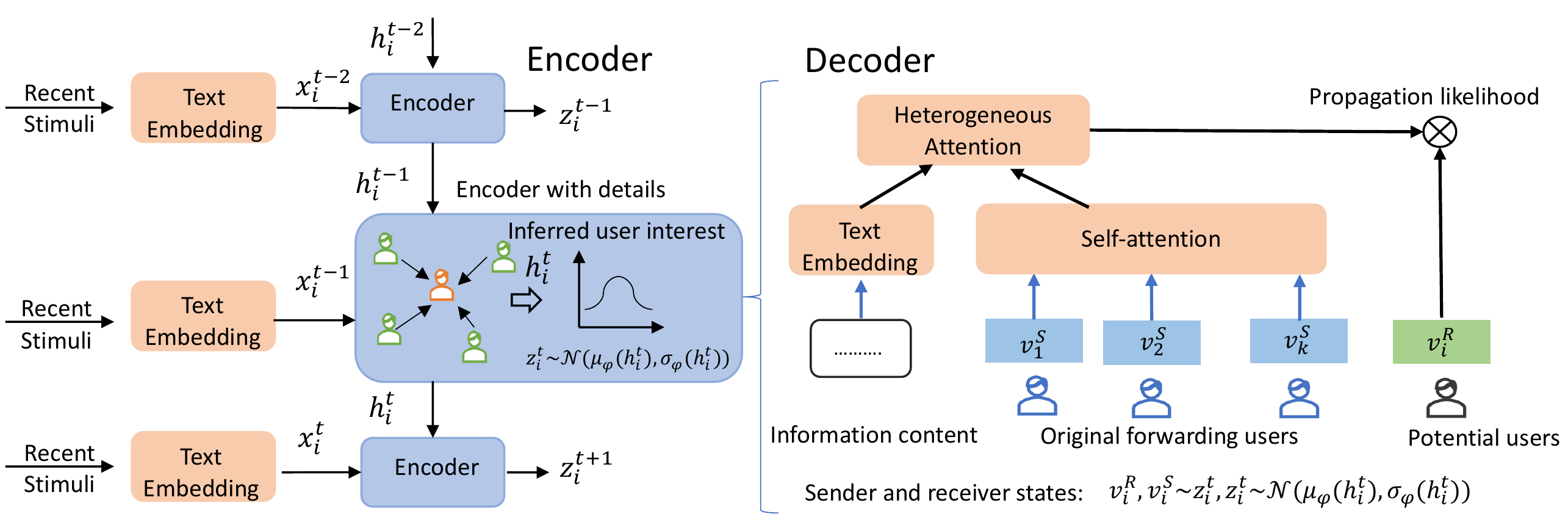}
    \caption{Overview of the DyDiff-VAE framework. The dynamic encoder updates the user interests a t new time step $t$ based on the recent stimuli and social influence. The dual attention decoder represents the information cascade based on content and user sequence, and estimates the propagation likelihood.}
    \label{fig:framework}
\end{figure*}

\subsection{Framework Overview}
At time step $t$, DyDiff-VAE estimates the propagation likelihood $p(u_i|I^{(t)}_j)$ for the potential users given an information cascade  $I^{(t)}_j = (c_j, \{u_{k}\}_{k = 1} ^{K})$ that consists of the cascade content $c_j$ and original forwarding user sequence $\{u_{k}\}_{k = 1} ^{K}$. The propagation likelihood is determined by the relevance between the information cascade and the potential users. We propose a variational framework for diffusion prediction, with the advance to infer evolving latent user interests $z^{(t)}$ along the time. The framework consists of two components:

\begin{enumerate}[leftmargin = 15 pt]
    \item \textbf{the dynamic encoder} (in Section~\ref{sec:encoder}) infers the posterior distribution of latent user interests $q_{\phi}(z^{(t)}|\cup_{n=1}^{n<t}\mathbb{I}^{(n)})$ at each time step by modeling recent stimuli and social influence from the historic information cascades;
    
    \item \textbf{the dual attention decoder} (in Section~\ref{sec:decoder}) estimates the propagation likelihood based on the inferred latent user interest $p_{\theta}(u_i|I_j^{(t)}, z_i^{(t)})$ by integrating both cascade content $c_j$ and original forwarding user sequence $\{u_{k}\}_{k = 1} ^{K}$.
\end{enumerate}

For each information cascade $I^{(t)}_j$, the framework optimizes the Evidence Lower BOund (ELBO)~\cite{VAE} $\mathcal{L}_{j}$ on the log likelihood of all users in the form of $\sum_{i}\log p(u_i|I_j^{(t)})$:

\begin{equation}
    \begin{split}
        & \sum_{i}\log p(u_i|I_j^{(t)})\\& \geq \mathcal{L}_{j}
         = \sum_{i} \mathbb{E}_{q_{\phi}(z_i^{(t)})} p_{\theta}(u_i|I_j^{(t)}, z_i^{(t)}) - KL(q_{\phi}(z^{(t)})\|p(z^{(t)})),
    \end{split}
    \label{eq:ELBO}
\end{equation}

\noindent
where $p(z_i^{(t)})$ denotes the prior distribution of $z_i^{(t)}$ approximated by unit normal distribution. For simplicity, we denote $q_{\phi}(z_i^{(t)})$ as the posterior distribution of user interests from our dynamic encoder in the remaining of our paper. Next, we will present the detailed designs of the encoder and decoder respectively.

\subsection{The Dynamic Encoder} \label{sec:encoder}
The dynamic encoder models the user interests evolution and infers the updated interests at new time steps for diffusion prediction. Two factors will influence the interests at new time steps: (1) \textbf{recent stimuli} which refer to the content users propagated recently, because users gain new information through such propagation behaviors; (2) \textbf{social influence} which describes the influence from the social neighbors (e.g., followees on Twitter) because the user interests can be influenced via the interactions with the social neighbors. The interests do not remain deterministic due to relatively nature of unpredictable and stochastic user behaviors. Considering that, at each time step, the dynamic encoder recursively models the user interests via a latent normal distribution based on the two factors from the last time step. For users at $t$-th time steps, let $x^{(t-1)}\in \mathbb{R}^{N\times d}$ encodes the information (s)he propagated at the last time step, $h^{(t)}\in \mathbb{R}^{N\times d}$ denotes the recurrent state, and $z^{(t)}\in \mathbb{R}^{N\times d}$ denotes the latent interest representation sampled from the normal distribution based on $h^{(t)}$. As shown in Figure~\ref{fig:framework}, the process to infer latent interests can be summarized as:
\begin{equation}
    h^{(t)} = f(g(h^{(t-1)}), x^{(t-1)}),
    \label{eq:RNN}
\end{equation}
\begin{equation}
    z^{(t)}_i \sim \mathcal{N}(\mu_{\phi}(h^{(t)}_i), \sigma_{\phi}(h^{(t)}_i)),
\end{equation}

\noindent
where $f(\cdot)$ is the recurrent function to update $h^{(t)}$ at each time step, $g(\cdot)$ integrates information from the social neighbors for each users, $\mu_{\phi}$ and $\sigma_{\phi}$ are the functions to characterize the normal distribution where $z^{(t)}_i$ is sampled. Next, we introduce the design of each part.

\subsubsection{\textbf{Recent stimuli $x^{(t-1)}$}} \label{sec:text}
At each time step, users would propagate some contents, and the propagated contents can influence the user interests at the next time step. To model the influence, we utilize $x^{(t-1)}\in \mathbb{R}^{N\times d}$ to encode the information from the propagated contents at the last time step, and it can be learned from any unsupervised/supervised text embedding algorithms~\cite{doc2vec,textCNN,bert} which map raw text to the representation vectors. The choice and comparison of the text embedding module are beyond the scope of this paper. We introduce the choice of this module in experiment part, and refer interested readers to explore various text embedding methods here.

\subsubsection{\textbf{Recurrent function $f(\cdot)$}}
As shown in Figure~\ref{fig:framework}, the encoder combines the social influence of $h^{(t-1)}$ and recent stimuli $x^{(t-1)}$, and updates $h^{(t)}$ for the new time step via Eq.~\ref{eq:RNN}. Intuitively, this combination should be automatically adjusted because these two factors influence the user interests differently and dynamically. Thus, we embed the graph convolutional layer into the gated recurrent unit (GRU) as the recurrent function $f(\cdot)$, where the graph convolutional layer naturally integrates information from the social neighbors via Graph Laplacian operator $\hat{D}^{-\frac{1}{2}} \hat{A} \hat{D}^{-\frac{1}{2}} h^{(t-1)}$, and GRU updates $h^{(t)}$ recursively. Specifically, at each time step:

\begin{equation}
    G_u = \sigma\left(\hat{D}^{-\frac{1}{2}} \hat{A} \hat{D}^{-\frac{1}{2}} h^{(t-1)} W_{hu} + W_{xu} x^{(t-1)}\right),
    \label{eq:gu}
\end{equation}
\begin{equation}
    G_r = \sigma\left(\hat{D}^{-\frac{1}{2}} \hat{A} \hat{D}^{-\frac{1}{2}} h^{(t-1)} W_{hr} + W_{xr} x^{(t-1)}\right),
    \label{eq:gr}
\end{equation}
\begin{equation}
    \tilde{h}^t = \tanh\left(\hat{D}^{-\frac{1}{2}} \hat{A} \hat{D}^{-\frac{1}{2}} (G_r \odot h^{(t-1)}) W_{hm} + W_{xm} x^{(t-1)}\right),
    \label{eq:htilde}
\end{equation}
\begin{equation}
    h^{(t)} = G_u \odot \tilde{h}^t + (1 -G_u) \odot h^{(t-1)},
    \label{eq:ht}
\end{equation}

\noindent
where $\sigma$ is sigmoid function, and $\odot$ denotes the Hadamard (element-wise) product, $\hat{D}^{-\frac{1}{2}} \hat{A} \hat{D}^{-\frac{1}{2}}$ is the normalized Graph Laplacian operator, $G_r$ is the reset gate, $G_u$ is the update gate, $W$ are trainable parameters. The reset gate $G_r$ determines how to form the combination $\tilde{h}^t$ from social influence and recent stimuli. And the update gate $G_u$ controls how these two factors influence the user original interests. One can also stack multiple graph convolutional layers in the GRU in order to consider the higher-order social influence in the social networks. 

\subsubsection{\textbf{Sampling from $\mathcal{N}(\mu_{\phi}(h^{(t)}_i), \sigma_{\phi}(h^{(t)}_i))$}}
To model the uncertainty of user's behaviors at each time step, we sample the current interest representations $z_i^{(t)}$ from the normal distributions $\mathcal{N}(\mu_{\phi}(h^{(t)}_i), \sigma_{\phi}(h^{(t)}_i))$, where $\mu_{\phi}(h^{(t)}_i)$ and $\sigma_{\phi}(h^{(t)}_i)$ are two linear layers learning mean and variance from $h^{(t)}_i$ respectively. With reparametrization tricks~\cite{VAE}, we first sample $\epsilon$ from $\mathcal{N}(0, I)$ and infer $z_i^{(t)}$ by:

\begin{equation}
    z_i^{(t)} = \mu_{\phi}(h^{(t)}_i) + \epsilon \odot \sigma_{\phi}(h^{(t)}_i)
    \label{eq:sampling}
\end{equation}

In the new step, we infer the latent interests $z_i^{(t)}$ for each user. Next, we introduce how to model the relevance between the information cascades and potential users based on $z_i^{(t)}$.

\subsection{The Dual Attention Decoder} \label{sec:decoder}
For a specific information cascade $I_j^{(t)}$ within time step $t$, the dual attention decoder estimates the propagation likelihood for each potential user $p(u_i|I_j^{(t)})$. Based on inferred user interests from the encoder, the decoder first represents the information cascade, then estimates the likelihood for each potential user. The key difference of our proposed dual attention decoder from other diffusion models~\cite{CYANRNN,SNIDSA,NDM,DeepDiffuse,Inf-VAE} is the consideration of both the cascade content $c_j$ and the user sequence $\{u_{k}\}_{k = 1} ^{K}$. Because in reality, cascade content also determines propagation behaviors of users, especially at the early stages the user sequences are short and reflect limited information. Next, we introduce how the dual attention decoder represents the information cascade and estimate the propagation likelihood in detail.

\subsubsection{\textbf{Propagation likelihood estimation}}
For a single information cascade occurring at time step $t$: $I^{(t)}_j = (c_j, \{u_{k}\}_{k = 1} ^{K})$, the propagation likelihood can be modeled as:
\begin{equation}
    p_{\theta}(u_i|I_j^{(t)}, z_i^{(t)}) = \sigma(<o_j, v_i^R>),
    \label{eq:likelihood}
\end{equation}

\noindent
where $\sigma(\cdot)$ is the sigmoid function, $<\cdot,\cdot>$ is the inner product, $o_j$ represents the observed information cascade $I^{(t)}_j$, $v_i^R$ represents the receiver state for potential user $u_i$. The same user appearing in observed information cascades and potential user set should have different states. Since in the former case the user would play as a sender who influence others, while in the latter one the user would play as a receiver who are influenced by the information cascade. This process is asymmetrical because senders are hardly influenced by receivers. For discussion convenience, we utilize the superscript $S$ to denote the vectors in the sender state and $R$ in the receiver state. Eq.~\ref{eq:SR} shows how we represent the sender state $v_i^S$ and receive state $v_i^R$ for $u_i$ based on $z_i^{(t)}$.

\begin{equation}
    v_i^S = W^S z_i^{(t)}, ~ v_i^R = W^R z_i^{(t)}
    \label{eq:SR}
\end{equation}

We omit superscript $t$ for $v_i^S$, $v_i^R$, because they both describe the states based on the same $z_i^{(t)}$ within $t$-th time step. Before introducing the techniques in the decoder, we want to highlight two challenges where we gain insights. 
\begin{enumerate}[leftmargin = 15pt]
    \item Some original forwarding users have noisy representations due to their sparse activities, and they may account for a large ratio due to the long tail distribution~\cite{longtail}. Therefore, it is critical to denoise and enhance their representations in order to characterize them more accurately.
    \item The information from the information content and the original forwarding users are heterogeneous, which reflects the information from a different perspective.
\end{enumerate}
 
Accordingly, we propose a dual attention decoder to address these challenges, where the first self-attention mechanism is designed to enhance and denoise the representations for original forwarding users, and the second heterogeneous attention mechanism is designed to aggregate information from both information content and user sequences.

\subsubsection{\textbf{Self-attention mechanism}}
We utilize the weighted sum of predecessors (together with the modeled user itself) as the denoised representation for each user, because users are more likely to be influenced by the predecessors within the original user sequence who they have interactions with. Thus we have:

\begin{equation}
    v_k = \sum_{i = 1}^k w_{ki} v_i^{S},
\end{equation}

\noindent   
where $v_k$ denotes the denoised representation for $u_k$. To estimate these weights, we consider two factors. First, as social influence brings interaction, the relevance between sender $u_i$ and receiver $u_j$, modeled by their inner product, may reflect influence. Second, the time elapsed between sender and receiver grows with decreasing influence. This is because the majority of social media users adopt more recent information, while often ignoring old and obsolete content~\cite{timeIC}. Therefore, we learn the weight $w_{ki}$ based on both users' inner product and temporal information, as follows:

\begin{equation}
    w_{ki} = \frac{\exp\left(<(\boldsymbol{v}_i^{S} + PE(i)), (\boldsymbol{v}_k^{R} + PE(k))>\right)}{\sum_{j = 1}^{k}\exp\left(<(\boldsymbol{v}_j^{S} + PE(j)), (\boldsymbol{v}_k^{R} + PE(k))>\right)},
\end{equation}

\begin{equation}
    PE(k)_{2 d} =\sin \left(\frac{k}{10000^{2 d / D}}\right), PE(k)_{2 d+1} =\cos \left(\frac{k}{ 10000^{2 d / D}}\right),
    \label{eq:PE}
\end{equation}

\noindent
where $<\cdot,\cdot>$ denotes the inner product of two vectors, $1 \leq d \leq D/2$ denotes the dimension of representations. We encode the relative position $k$ through positional-encodings $PE(k)$~\cite{transformer} as shown in Eq.~\eqref{eq:PE}.

\subsubsection{\textbf{Heterogeneous-attention mechanism}}
To integrate heterogeneous information from information cascade $I^{(t)}_j$, we first represent the information content $c_j$ via the same text embedding module mentioned in Section~\ref{sec:text}, and then project it into the same space where the users are represented through one linear layer. Let $v_c$ denote the learned vector. Then we propose an attention mechanism to integrate $(v_c, \{v_{k}\}_{k = 1} ^{K})$ into the cascade representation $o_j$:
\begin{equation}
    o_j = \alpha_c v_c + \sum_{k=1}^K \alpha_k v_{k},
    \label{eq:cascadeembedding}
\end{equation}
where $\alpha_c$ and $\alpha_k$ are attention weights to combine information content and user sequence:
\begin{equation}
    \begin{split}
        \alpha_c &= \frac{\exp(<v_c, v_c>)}{\exp(<v_c, v_c>) + \sum_{k=1}^K\exp(<v_c, v_k>)}, \\
        \alpha_k &= \frac{\exp(<v_k, v_c>)}{\exp(<v_c, v_c>) + \sum_{k=1}^K\exp(<v_c, v_k>)}
    \end{split}
\end{equation}

\begin{algorithm}[t]
\caption{DyDiff-VAE training procedure.}
\label{al:training}
\KwIn{Social Network $\mathcal{G}$, Training cascades $\bigcup_{n=1}^{n<t}\mathbb{I}^{t}$.}
\KwOut{Model parameters $\phi$ and $\theta$.}
\While{model not converged}{
    \For{Each time step $t$ during training}{
        (\textit{For the dynamic encoder}:)\\
        Learn user recent stimuli $x^{(t-1)}$ by text embedding on the content propagated at time step $t-1$;\\
        Update new interests $z^{(t)}$ according to Eq.~\ref{eq:gu} - Eq.~\ref{eq:sampling}; \\
        (\textit{For the dual attention decoder}:)\\
        \For{Each information cascade at $t$ time step}{
            Based on $z^{(t)}$, estimate propagation likelihood according to Eq~\ref{eq:cascadeembedding} and Eq~\ref{eq:likelihood};\\
            Calculate ranking loss according to Eq.~\ref{eq:diffloss};\\
        }
    }
    Update the parameters $\phi$ and $\theta$ by optimizing variational ranking loss in Eq.~\ref{eq:obj};
}
\end{algorithm}

\subsection{Optimization}
DyDiff-VAE optimizes the variational lower bound on the log likelihood of all observed information cascades. For one single cascade, the objective is shown in Eq~\ref{eq:ELBO}, where the first term is the reconstruction loss to estimate the users' propagation behaviors. For implementation, instead of estimating the likelihood expectation (via Monte Carlo sampling) , we use the ranking loss to reconstruct the observed cascades. For one thing, it is hard to correctly observe all forwarding users in datasets, so the estimated expectation would be biased due to noise. Also, in real world applications, (e.g., information recommendation), the ranking results are more practical than the specific likelihood. Thus, we directly optimize the weighted approximate-rank pairwise loss (WARP)~\cite{warp} and expect observed forwarding users (positive samples) to be ranked higher than the other users (negative samples). Specifically, for information cascade $I_j^{(t)}$, the ranking loss is defined as:

\begin{equation}
    \begin{split}
        \mathcal{L}_{j}^{DIFF} &= \sum_{u_{i^{+}}} \frac{\sum_{u_{i^{-}}} L(rank(u_{i^{+}})) \cdot \mathcal{L}_{j}^{pair}(u_{i^{+}}, u_{i^{-}})}{rank(u_{i^{+}})},\\
        \mathcal{L}_{j}^{pair}(u_{i^{+}}, u_{i^{-}}) &= \max(0, \lambda_{m} - p_{\theta}(u_i^{+}|I_j^{(t)}, z_i^{(t)}) + p_{\theta}(u_i^{-}|I_j^{(t)}, z_i^{(t)})),
    \end{split}
    \label{eq:diffloss}
\end{equation}

\noindent
where $u_{i^{+}}$ denotes the positive samples, $u_{i^{-}}$ denotes the negative samples, $L(K) = \sum_{k = 1}^{K}1/k$. For each observed positive sample $u_{i^{+}}$, we expect the likelihood $p_{\theta}(u_i^{+}|I_j^{(t)}, z_i^{(t)})$ should be larger than that of any negative samples by the margin $\lambda_m$., i.e., $\mathcal{L}_{j}^{pair}(u_{i^{+}}, u_{i^{-}}) = 0$. Otherwise, we penalize each pair of $(u_i^{+}, u_i^{-})$ because of the wrong ranking, and $L(rank(u_{i^{+}}))$ is the penalty weight.

Thus, our overall objective is to optimize the variational ranking loss in Eq~\ref{eq:obj}. 
\begin{equation}
    \begin{split}
        \mathcal{L} &= \underbrace{\sum_{j}\mathcal{L}_{j}^{DIFF}}_{\text{ranking loss}} + \underbrace{\beta \cdot \sum_{t} \left( KL(q_{\phi}(z^{(t)})\|p(z^{(t)})) \right)}_{\text{KL divergence for all time steps}} \\
        &+ \underbrace{\lambda_{1} \cdot \|\phi\|_2 + \lambda_{2} \cdot \|\theta\|_2}_{\text{regularization}},
    \end{split}
    \label{eq:obj}
\end{equation}

\noindent
where $\beta$ is the weight of KL divergence, $\lambda_1$ and $\lambda_2$ are the regularization weights for encoder and decoder parameters. Algorithm~\ref{al:training} summarizes the training procedure of DyDiff-VAE.

\subsection{Diffusion Prediction and Ranking}
We optimize the model parameters on information cascades at previous time steps, i.e., $\bigcup_{n=1}^{n<t}\mathbb{I}^{n}$. After convergence, DyDiff-VAE can be applied to predict the propagation likelihood and the corresponding user rankings for cascades at the new time step $t$. Given the observed $I^{(t)}_j = (c_j, \{u_{k}\}_{k = 1} ^{K})$ at the new time step, DyDiff-VAE first infers the user new interests according to Eq.~\ref{eq:sampling}, then estimates likelihoods for all potential users according to Eq.~\ref{eq:likelihood}, which induces the user rankings. 

\section{experiment}
\label{sec:experiment}
In this section, we present and discuss our experimental results as well as ablation studies and parameter analysis on real-world datasets.

\subsection{Datasets}
Since existing datasets lack the content information, we collect the following four datasets from Twitter and Youtube to evaluate our proposed model. We collect data related to {\em Election\/}  \footnote{\url{https://en.wikipedia.org/wiki/2018_Venezuelan_presidential_election}}, {\em Venezuela\/} \footnote{\url{https://en.wikipedia.org/wiki/Venezuela}}, {\em White Helmets \/} \footnote{\url{https://en.wikipedia.org/wiki/White_Helmets_(Syrian_Civil_War)}} to form three Twitter datasets and one YouTube data set. On Twitter, each retweet sequence of a content item constitutes a diffusion cascade, while the explicit social network consists of follower-followee links. Following~\cite{DeepDiffuse,SNIDSA,Inf-VAE}, we removed tweets with non-English content and preserved users with more than $10$ records and cascades longer than $10$. On YouTube, user interactions constitute the social networks, and a cascade corresponds to the comment sequence of a video. We collected data related to Venezuela. Based on the raw data, We extract the video descriptions as the propagated cascade contents, and then preserve users with more than $5$ records and cascades longer than $10$.

\begin{table}[t]
\small
\centering
\caption{Statistics of datasets used in our experiments.}
\label{tb:dataset}
\begin{tabular}{c|ccc|c}
\hline
\textbf{Platform} & \multicolumn{3}{c|}{\textbf{Twitter}} & \textbf{Youtube} \\ \hline
\textbf{Topic} & \textbf{Election} & \textbf{Venezuela} & \textbf{While Helmet} & \textbf{Venezuela} \\ \hline
\textbf{\# Users} & 5,948 & 5,069 & 2,373 & 6,393 \\
\textbf{\# Links} & 30,762 & 34,622 & 15,706 & 4,738 \\
\textbf{\# Cascades} & 1,514 & 2,980 & 623 & 321 \\
\textbf{Avg. len} & 40.6 & 38.2 & 117.2 & 60.9 \\
\textbf{Start Date} & 2018-12 & 2018-12 & 2018-7 & 2018-12 \\
\textbf{End Date} & 2019-02 & 2019-03 & 2019-04 & 2019-03 \\ \hline
\end{tabular}
\end{table}

    
    
    

Dataset statistics are provided in Table~\ref{tb:dataset}. All referred IDs have been anonymized. For each dataset, we uniformly separate the time span into $6$ discrete time steps and utilize cascades in the first five time steps as Training set. Val/Test are split randomly on the information cascades on $6$-th time step, with ratio $1:3$. Following \cite{DeepDiffuse,TopoLSTM,Inf-VAE}, we remove users who do not exist in Training set from Val/Test set, for we have no information of these users. We also try different time segmentation strategies, where our method consistently outperforms others. We leave a systematic study for optimal segmentation as our future work.

\begin{table*}[t]
\small
\centering
\caption{$MAP@K$ for diffusion prediction on 4 datasets.}
\label{tb:map}
\begin{tabular}{c|ccccccccc|ccc}
\hline
\textbf{Platform} & \multicolumn{9}{c|}{\textbf{Twitter}} & \multicolumn{3}{c}{\textbf{Youtube}} \\ \hline
\textbf{Dataset} & \multicolumn{3}{c|}{\textbf{Election}} & \multicolumn{3}{c}{\textbf{Venezuela}} & \multicolumn{3}{c|}{\textbf{White Helmet}} & \multicolumn{3}{c}{\textbf{Venezuela}} \\ \hline
\textbf{MAP} & \textbf{@10} & \textbf{@50} & \multicolumn{1}{c|}{\textbf{@100}} & \textbf{@10} & \textbf{@50} & \multicolumn{1}{c|}{\textbf{@100}} & \textbf{@10} & \textbf{@50} & \textbf{@100} & \textbf{@10} & \textbf{@50} & \textbf{@100} \\ \hline
\textbf{Random} & 0.0011 & 0.0010 & \multicolumn{1}{c|}{0.0009} & 0.0010 & 0.0009 & \multicolumn{1}{c|}{0.0010} & 0.0034 & 0.0048 & 0.0028 & 0.0019 & 0.0008 & 0.0010 \\ \hline
\textbf{Node2vec} & 0.0102 & 0.0110 & \multicolumn{1}{c|}{0.0122} & 0.0105 & 0.0147 & \multicolumn{1}{c|}{0.0168} & 0.0095 & 0.0136 & 0.0155 & 0.0067 & 0.0053 & 0.0041 \\
\textbf{GAE} & 0.0258 & 0.0252 & \multicolumn{1}{c|}{0.0304} & 0.0229 & 0.0250 & \multicolumn{1}{c|}{0.0292} & 0.0326 & 0.0267 & 0.0315 & 0.0092 & 0.0063 & 0.0058 \\
\textbf{VGRNN} & 0.0137 & 0.0108 & \multicolumn{1}{c|}{0.0186} & 0.012 & 0.0109 & \multicolumn{1}{c|}{0.0173} & 0.0158 & 0.0181 & 0.0177 & 0.0083 & 0.0073 & 0.0063 \\ \hline
\textbf{GraphRec} & 0.0224 & 0.0241 & \multicolumn{1}{c|}{0.0255} & 0.0298 & 0.0280 & \multicolumn{1}{c|}{0.0300} & 0.0233 & 0.0171 & 0.0177 & 0.0023 & 0.0024 & 0.0021 \\ \hline
\textbf{DeepDiffuse} & 0.0276 & 0.0264 & \multicolumn{1}{c|}{0.0315} & 0.0171 & 0.0208 & \multicolumn{1}{c|}{0.0232} & 0.0351 & 0.0252 & 0.0291 & 0.0120 & 0.0106 & 0.0119 \\
\textbf{SNIDSA} & 0.0299 & 0.0296 & \multicolumn{1}{c|}{0.0353} & 0.0419 & 0.0376 & \multicolumn{1}{c|}{0.0406} & 0.3045 & 0.0274 & 0.0277 & 0.0131 & 0.0111 & 0.0124 \\
\textbf{Topo-LSTM} & 0.0307 & 0.0347 & \multicolumn{1}{c|}{0.0397} & 0.0325 & 0.0361 & \multicolumn{1}{c|}{0.0393} & 0.0293 & 0.0262 & 0.0291 & 0.0081 & 0.0067 & 0.0071 \\
\textbf{Inf-VAE} & 0.0423 & 0.0406 & \multicolumn{1}{c|}{0.0454} & 0.0393 & 0.0401 & \multicolumn{1}{c|}{0.0435} & 0.0330 & 0.0257 & 0.0289 & 0.0116 & 0.0102 & 0.0094 \\
\textbf{Inf-VAE+Content} & 0.0428 & 0.0378 & \multicolumn{1}{c|}{0.0432} & 0.0425 & 0.0416 & \multicolumn{1}{c|}{0.0451} & 0.0426 & 0.0275 & 0.0298 & 0.0099 & 0.0076 & 0.0063 \\ \hline
\textbf{DyDiff-VAE} & \textbf{0.0724*} & \textbf{0.0698*} & \multicolumn{1}{c|}{\textbf{0.0764*}} & \textbf{0.0602*} & \textbf{0.0601*} & \multicolumn{1}{c|}{\textbf{0.0626*}} & \textbf{0.0570*} & \textbf{0.0577*} & \textbf{0.0502*} & \textbf{0.0169*} & \textbf{0.0134*} & \textbf{0.0141*} \\ \hline
\textbf{Imprv.} & \multicolumn{1}{l}{\textit{+69.2\%}} & \multicolumn{1}{l}{\textit{+71.9\%}} & \multicolumn{1}{l|}{\textit{+68.3\%}} & \multicolumn{1}{l}{\textit{+41.6\%}} & \multicolumn{1}{l}{\textit{+44.5\%}} & \multicolumn{1}{l|}{\textit{+38.8\%}} & \multicolumn{1}{l}{\textit{+33.8\%}} & \multicolumn{1}{l}{\textit{+109.8\%}} & \multicolumn{1}{l|}{\textit{+59.4\%}} & \multicolumn{1}{l}{\textit{+29.0\%}} & \multicolumn{1}{l}{\textit{+20.7\%}} & \multicolumn{1}{l}{\textit{+13.7\%}} \\ \hline
\end{tabular}
\end{table*}

\begin{table*}[t]
\small
\centering
\caption{$Recall@K$ for diffusion prediction on 4 datasets.}
\label{tb:recall}
\begin{tabular}{c|ccccccccc|ccc}
\hline
\textbf{Platform} & \multicolumn{9}{c|}{\textbf{Twitter}} & \multicolumn{3}{c}{\textbf{Youtube}} \\ \hline
\textbf{Dataset} & \multicolumn{3}{c|}{\textbf{Election}} & \multicolumn{3}{c}{\textbf{Venezuela}} & \multicolumn{3}{c|}{\textbf{White Helmet}} & \multicolumn{3}{c}{\textbf{Venezuela}} \\ \hline
\textbf{Recall} & \textbf{@10} & \textbf{@50} & \multicolumn{1}{c|}{\textbf{@100}} & \textbf{@10} & \textbf{@50} & \multicolumn{1}{c|}{\textbf{@100}} & \textbf{@10} & \textbf{@50} & \textbf{@100} & \textbf{@10} & \textbf{@50} & \textbf{@100} \\ \hline
\textbf{Random} & 0.0011 & 0.0100 & \multicolumn{1}{c|}{0.0193} & 0.0010 & 0.0085 & \multicolumn{1}{c|}{0.0150} & 0.0034 & 0.0185 & 0.0323 & 0.0019 & 0.0008 & 0.0010 \\ \hline
\textbf{Node2vec} & 0.0152 & 0.0723 & \multicolumn{1}{c|}{0.1123} & 0.0252 & 0.0913 & \multicolumn{1}{c|}{0.1420} & 0.0189 & 0.0713 & 0.1208 & 0.0066 & 0.0243 & 0.0346 \\
\textbf{GAE} & 0.0304 & 0.1200 & \multicolumn{1}{c|}{0.2260} & 0.0389 & 0.1417 & \multicolumn{1}{c|}{0.2418} & 0.0376 & 0.1298 & 0.2345 & 0.0061 & 0.0173 & 0.0471 \\
\textbf{VGRNN} & 0.0176 & 0.0548 & \multicolumn{1}{c|}{0.1185} & 0.0199 & 0.1019 & \multicolumn{1}{c|}{0.1474} & 0.0253 & 0.0874 & 0.1327 & 0.0089 & 0.0218 & 0.0386 \\ \hline
\textbf{GraphRec} & 0.0147 & 0.0837 & \multicolumn{1}{c|}{0.1429} & 0.0334 & 0.1261 & \multicolumn{1}{c|}{0.1915} & 0.0303 & 0.1201 & 0.1977 & 0.0027 & 0.0044 & 0.0115 \\ \hline
\textbf{DeepDiffuse} & 0.0382 & 0.1318 & \multicolumn{1}{c|}{0.2172} & 0.0329 & 0.1337 & \multicolumn{1}{c|}{0.1966} & 0.0499 & 0.1363 & 0.2076 & 0.0145 & 0.0296 & 0.0421 \\
\textbf{SNIDSA} & 0.0452 & 0.1232 & \multicolumn{1}{c|}{0.2253} & 0.0463 & 0.1464 & \multicolumn{1}{c|}{0.2199} & 0.0457 & 0.1374 & 0.2138 & 0.0132 & 0.0261 & 0.0410 \\
\textbf{Topo-LSTM} & 0.0506 & 0.1727 & \multicolumn{1}{c|}{0.2568} & 0.0447 & 0.1419 & \multicolumn{1}{c|}{0.2114} & 0.0304 & 0.1046 & 0.1965 & 0.0081 & 0.0217 & 0.0328 \\
\textbf{Inf-VAE} & 0.0462 & 0.1642 & \multicolumn{1}{c|}{0.2304} & 0.0610 & 0.1586 & \multicolumn{1}{c|}{0.2390} & 0.0414 & 0.1377 & 0.2041 & 0.0125 & 0.0259 & 0.0409 \\
\textbf{Inf-VAE+Content} & 0.0529 & 0.1599 & \multicolumn{1}{c|}{0.2510} & 0.0623 & 0.1750 & \multicolumn{1}{c|}{0.2541} & 0.0441 & 0.1295 & 0.2183 & 0.0106 & 0.0266 & 0.0327 \\ \hline
\textbf{DyDiff-VAE} & \textbf{0.0862*} & \textbf{0.2377*} & \multicolumn{1}{c|}{\textbf{0.3376*}} & \textbf{0.0944*} & \textbf{0.2204*} & \multicolumn{1}{c|}{\textbf{0.2892*}} & \textbf{0.0809*} & \textbf{0.1850*} & \textbf{0.2695*} & \textbf{0.0188*} & \textbf{0.0371*} & \textbf{0.0561*} \\ \hline
\textbf{Imprv.} & \textit{+62.9\%} & \textit{+37.6\%} & \multicolumn{1}{c|}{\textit{+31.5\%}} & \textit{+47.4\%} & \textit{+25.9\%} & \multicolumn{1}{c|}{\textit{+13.8\%}} & \textit{+62.1\%} & \textit{+48.7\%} & \textit{+34.4\%} & \textit{+29.7\%} & \textit{+25.3\%} & \textit{+19.1\%} \\ \hline
\end{tabular}
\end{table*}

\subsection{Baselines}
We compare our model against state-of-the-art methods from related areas including graph representation learning (dynamic/static), social recommendation, and diffusion prediction.
\begin{itemize}[leftmargin = 10pt]
    \item \textbf{Node2vec}~\cite{node2vec}: a framework that learns low-dimensional representations for nodes in a network by optimizing a neighborhood preserving objective.
    \item \textbf{GAE}~\cite{gae}: a graph autoencoder framework consists of graph convolutional layers to represent the network and node attributes, and an inner product decoder to predict the link existence. 
    \item \textbf{VGRNN}~\cite{VGRNN}: a variational autoencoder framework for dynamic graph learning, which represents the dynamic adjcency matrix and node attributes, and predicts the link existence.
    \item \textbf{GraphRec}~\cite{GraphRec}: a novel graph neural network framework for social recommendations, which coherently models interactions and opinions in the user-item graph.
    \item \textbf{DeepDiffuse}~\cite{DeepDiffuse}: an attention-based RNN that operates on the subsequence of previously influenced users to predict diffusion, where the subsequence is extracted by another modified LSTM.
    \item \textbf{SNIDSA}~\cite{SNIDSA}: a novel sequential neural network with structure attention to model both sequential nature of an information diffusion process and structural characteristics of user connection graph for diffusion prediction.
    \item \textbf{Topo-LSTM}~\cite{TopoLSTM}: a recurrent model that exploits the local propagation structure during the diffusion process through a dynamic tree-based LSTM.
    \item \textbf{Inf-VAE}~\cite{Inf-VAE}: a variational framework that combines a variation graph autoencoder and a co-attention network to jointly model social relations and sequence information.
\end{itemize}

\subsection{Experimental Setup}
Following~\cite{DeepDiffuse,CYANRNN,Inf-VAE,SNIDSA}, we evaluate our diffusion model in a retrieval setting. We rank potential users based on their propagation likelihood and utilize $MAP@K$ and $Recall@K$ as the metrics to evaluate the predicted rankings. During the evaluation, unless specifically mentioned, we keep the first half of forwarding users observed in a ground truth cascade and predict the other half of forwarding users. This ratio is defined as \textit{seed set percentage}, and we will also report the performance comparison under various seed set percentages.

To train our model, we fix the dimension of our representation vectors as $128$, regularization weight of model parameters $\lambda_{1}$ and $\lambda_{2}$ as $0.5$, the KL divergence weight $\beta$ as $10$. Then, we tune learning rate and $\lambda_{m}$, by evaluating $MAP@10$ on the Val set. We choose the unsupervised Doc2Vec~\cite{doc2vec} as the text embedding module for simplicity. For Node2vec, GAE, and VGRNN, the outcomes are the user representations, we use the average aggregation function to represent the cascades, which has the best performance compared with Sum and Earliest (root representations) functions. For GraphRec, since it is designed to predict the rating scores among each user-item pair in a standard recommendation setting, we set the rating score as $1$ if a user forwards the cascade, otherwise $0$. We use the ranking induced by rating score for diffusion prediction. For other diffusion baselines, as stated in~\cite{Inf-VAE}, we use the ranking induced by propagation likelihood for diffusion prediction.  

\subsection{Experimental Results}
\subsubsection{\textbf{Overall Results}}
Table~\ref{tb:map} and Table~\ref{tb:recall} show the performance comparisons, where the last row presents the relative improvement over the best baseline~\footnote{All results are averaged over $5$ independent runs with different random seeds. $^*$ indicates the statistically significant improvements over the best baseline, with p-value smaller than $10^{-6}$.}. Node2vec considers the social structure information, GAE and VGRNN consider both structure information as well as user interests learnt from contents, VGRNN models the evolution of user interests, GraphRec jointly models the user-cascade interactions and cascade contents in the recommendation framework, and the other diffusion baselines study the sequential information of the cascade of original forwarding users. 

DyDiff-VAE significantly outperforms all the baselines on both $MAP$ and $Recall$ metrics, meaning that more forwarding users are found and they rank higher in the predicted list. The diffusion prediction is indeed a hard task, which is indicated by the extremely low scores from {\em Random \/}results, where the {\em Random \/}result could be $0.50$ in some classification tasks. However, our model is able to achieve superior result. For example on Twitter/Election, we aims to retrieve $\sim20$ users out of $\sim5$k users based on initial $\sim20$ users. A $Recall@100$ of $0.34$ is quite impressive, considering the absence of explicit propagation tree topology and the noise in the real-world diffusion processes. 

Also, we find existing diffusion methods are not effective at modeling cascade content and user interests. We modified Inf-VAE to consider such information as its GCN-based encoder can be easily generalized to model user attributes. We call the modified baseline as \textit{Inf-VAE+Content}. Results do not show a consistent improvement over the original Inf-VAE. It is worth noting that the dataset characteristics influence prediction performance. It is easier to predict on Twitter/Election dataset consisting of the polarized posts because users having different stances behave quite differently. For the Youtube dataset, more flexible social interactions and more diverse cascade content make it less predictable. 

\begin{figure}[t]
    \centering
    \begin{subfigure}[b]{0.49\linewidth}
         \centering
         \includegraphics[width = \linewidth]{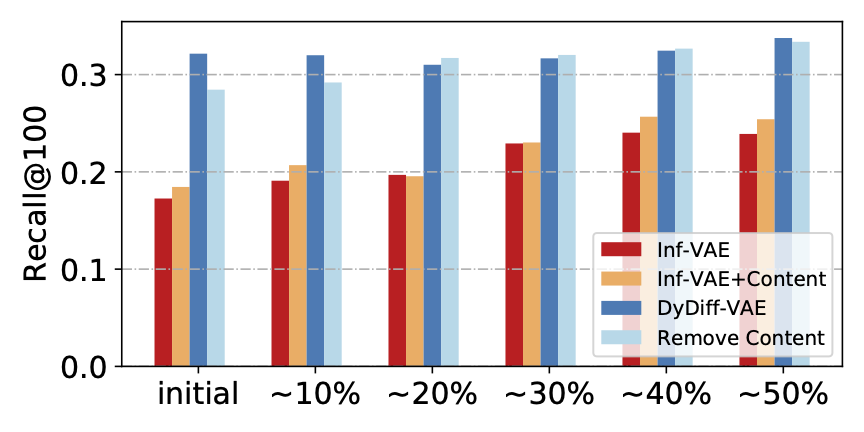}
         \caption{Twitter/Election}
         \label{fig:pap1}
    \end{subfigure}
    \hfill
    \begin{subfigure}[b]{0.49\linewidth}
         \centering
         \includegraphics[width = \linewidth]{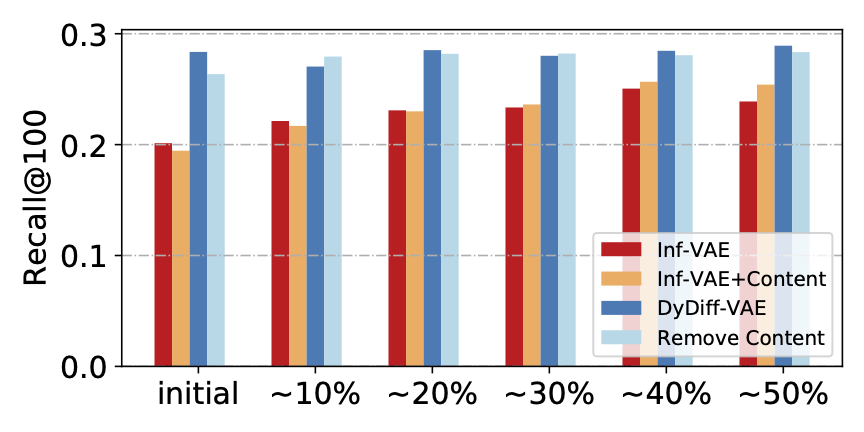}
         \caption{Twitter/Venezuela}
         \label{fig:pap2}
    \end{subfigure}
    \caption{Recall@100 under various seed set percentage.}
    \label{fig:ratio}
\end{figure}

\subsubsection{\textbf{Results under various seed set percentages}} \label{sec:content}
A desirable property of diffusion model is that the it has robust performance when observing different ratio of original forwarding users. We vary the seed set percentage from \textit{initial} (meaning that we only observe the initial posts) to $0.5$. We use dataset \textit{Twitter/Election} and \textit{Twitter/Venezuela} for this study, for we require a sizable number of test examples to obtain unbiased estimates. The results are shown in Figure~\ref{fig:ratio}. Recall increases with seed set percentage as expected. We compare our results with two strong baselines. It is interesting to find that under \textit{initial} situation, their performance drop significantly, while DyDiff-VAE still achieves competitive performance. To dig into the reason, we show the result after removing content information (by removing the heterogeneous attention layer). We observe a performance drop in the initial situation, because in the beginning, the user sequences are too short to provide sufficient information. As the ratio increases, the performances tend to be close,  because information content does not provide more information after observing the first half of the user sequence. 

\subsection{Ablation Study and Analysis}
\subsubsection{\textbf{Ablation Study}}
We conduct an ablation study to analyze our design choices.
We first examine the effectiveness of the dynamic encoder:
\begin{enumerate}[leftmargin = 15pt]
    \item \textbf{Static encoder}. The user interests are assumed static, and the dynamic encoder becomes one-layer GCN.
    \item \textbf{Remove Conv.}. We eliminate the social influence, so the user interests will be independent of other users.
\end{enumerate}

\begin{table}[t]
\small
\centering
\caption{Ablation study on framework design.}
\label{tb:ablation}
\begin{tabular}{l|cc|cc}
\hline
\textbf{Dataset} & \multicolumn{2}{c|}{\textbf{Twitter/Election}} & \multicolumn{2}{c}{\textbf{Twitter/Venezuela}} \\ \hline
\textbf{Metrics (@100)} & \textbf{MAP} & \textbf{Recall} & \textbf{MAP} & \textbf{Recall} \\ \hline
\textbf{(1) Static encoder} & 0.0530 & 0.2660 & 0.0516 & 0.2579 \\
\textbf{(2) Remove Conv.} & 0.0650 & 0.3007 & 0.0485 & 0.2650 \\ \hline
\textbf{(3) Remove 1st attn} & 0.0466 & 0.2572 & 0.0327 & 0.2360 \\
\textbf{(4) Remove 2nd attn} & 0.0738 & 0.3337 & 0.0561 & 0.2734 \\
\textbf{(5) $W^S = W^R$} & 0.0693 & \multicolumn{1}{c|}{0.3199} & \multicolumn{1}{c}{0.0591} & \multicolumn{1}{c}{0.2710} \\ \hline
\textbf{(6) DyDiff-AE} & \multicolumn{1}{c}{0.0746} & \multicolumn{1}{c|}{0.3124} & \multicolumn{1}{c}{0.0603} & \multicolumn{1}{c}{0.2630} \\\hline
\textbf{DyDiff-VAE} & \textbf{0.0764} & \textbf{0.3376} & \textbf{0.0626} & \textbf{0.2892} \\ \hline
\end{tabular}
\end{table}

Replacing the dynamic encoder as a static one will result in a significant performance drop. It proves the importance of modeling user dynamic interests on diffusion prediction. Further, while modeling such evolution, the worse performance after removing the graph convolutional layer illustrates the necessity to consider the social influence from other users. Next, we analyze the effectiveness of the dual attentive decoder by the following ablations:

\begin{enumerate}[leftmargin = 15pt,resume]
    \item \textbf{Remove 1st attn}. The cascade representations are directly learned on the user interests from encoder, without any denoising and enhancement.
    \item  \textbf{Remove 2nd attn}. Only consider the user sequence without modeling the information content. The cascade representation is the average on the denoised user representations. 
    \item $W^S = W^R$. Ignore the difference between the sender state and receiver state.
\end{enumerate}
In our dual attentive decoder, we utilize a self-attention layer to learn the local influence and denoise the user interest representations, especially for inactive users. Removing such method  (3) results in a significant performance drop. Notably, replacing the heterogenous attention layer with a mean aggregation on user sequence leads to a close performance. Because, as stated in Section~\ref{sec:content} the information content does not provide more information after observing the first half of the user sequence. Moreover, overlooking asymmetric social influence (5) slightly deteriorates results, which indicates the existence of asymmetric influence among users. Finally, we analyze the effectiveness of the variational framework by the following ablations:
\begin{enumerate}[leftmargin = 15pt,resume]
    \item \textbf{DyDiff-AE}. We replace the variational framework with a autoencoder framework, and the user interests are deterministics.
\end{enumerate}
As is expected, DyDiff-AE results in a worse performance, since it overlooks the uncertainty of the user interests.
 

\subsubsection{\textbf{Impact of graph convolution}}
Social influence has been proved necessary to model the user dynamic interests. We further study whether stacking more graph convolutional layers, in order to integrate higher-order social influence, can help for diffusion prediction. As shown in Figure~\ref{fig:gcn}, adding one more layer increase the $Recall@100$ on Twitter/Election dataset but have a negative impact on Twitter/Venezuela dataset. This is because, in part, that the social network of Twitter/Venezuela is far denser, and multiple convolutional layers may cause the over-smoothing issue~\cite{oversmooth,oversmooth2}.  

\subsubsection{\textbf{Impact of $\lambda_m$}}
\begin{figure}[t]
    \centering
    \begin{subfigure}[b]{0.38\linewidth}
         \centering
         \includegraphics[width = \linewidth]{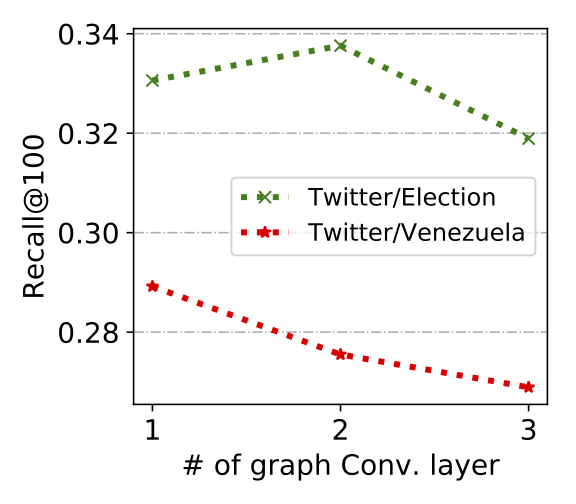}
         \caption{Impact of Conv. Layers}
         \label{fig:gcn}
    \end{subfigure}
    \hfill
    \begin{subfigure}[b]{0.59\linewidth}
         \centering
         \includegraphics[width = \linewidth]{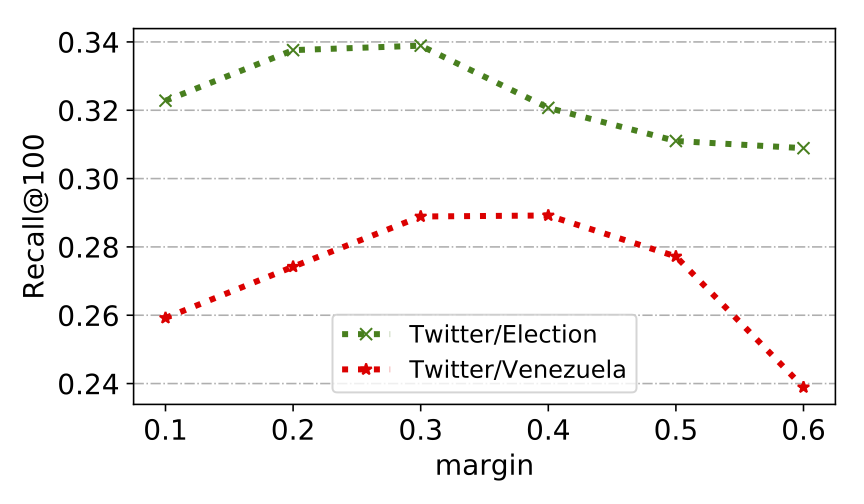}
         \caption{Impact of $\lambda_m$}
         \label{fig:lambda}
    \end{subfigure}
    \caption{Parameter analysis evaluated by $Recall@100$.}
    \label{fig:impact}
\end{figure}

We study the impact $\lambda_{m}$, as shown in Figure~\ref{fig:lambda}. $\lambda_{m}$ is the marginal value to penalize negative users ranking higher. Large $\lambda_{m}$ can hurt the performance, for it ignores some relevance, though is not captured in the dataset, between the cascades and negative users.

\subsubsection{\textbf{Efficiency Analysis}}
\begin{figure}[t]
    \centering
    \includegraphics[width = 0.8\linewidth]{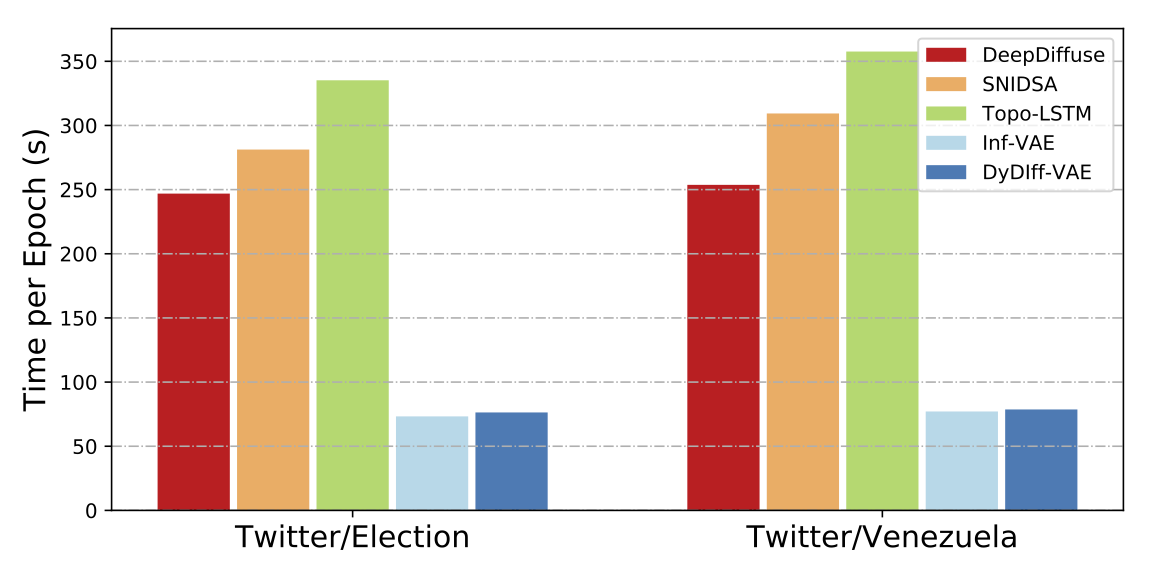}
    \caption{Running time comparison, unit: second (s). DyDiff-VAE is faster than recurrent models (DeepDiffuse, SNIDSA, Topo-LSTM), and close to attention model (Inf-VAE).}
    \label{fig:time}
\end{figure}

As mentioned earlier, efficiency is the bottleneck of sequence modeling for large-scale applications. We run diffusion baselines and DyDiff-VAE on the same workstation equipped with an Intel i9-9960X processor, 64GB memory, and one NVIDIA RTX 2080 Ti GPU, and compare execution time per epoch (average on 20) on two large datasets.  Each model is trained with one GPU on the device.  Figure~\ref{fig:time} shows the comparison results. DyDiff-VAE is faster than the other RNN-based baselines. And it achieves the competitive efficiency as Inf-VAE with consistently superior performance.

\section{conclusion}
\label{sec:conclusion}

In this paper, we proposed a novel dynamic variational framework that jointly considers dynamic user interests and information content for diffusion prediction. We propose a dynamic encoder to infer the dynamic user interests from historical behaviors, and a dual attentive decoder to capture the information from both the information cascade and the sequence of original users to estimate the propagation likelihood. Our experimental results on four real-world datasets demonstrate that the proposed model outperforms other compared methods. 

The work has several shortcomings that offer avenues for future extensions. First, the algorithms are described at a fixed point in time when some data have been collected and the future is to be predicted. In reality, incremental algorithms are needed where new data arrive over time. Second, we do not take into account the fact that data sets are invariably incomplete. Thus, it is likely that future cascade participants will include users never before seen in the collected data. In the future, we plan to extend DyDiff-VAE to generate realistic user profiles (in terms of social relations and interests) from the learned latent distribution for unseen users. 

\section*{Acknowledgements}
This work was funded in part by DARPA under award W911NF-17-C-0099, and by DoD Basic Research Office under award HQ00342110002.

\newpage
\balance
\bibliographystyle{plain}
\bibliography{main.bib}
\end{document}